\documentclass[9pt,conference]{IEEEtran}
\IEEEoverridecommandlockouts
\usepackage{spconf,amsmath,graphicx,hyperref}

\usepackage[utf8]{inputenc} 
\usepackage[T1]{fontenc}    
\usepackage{hyperref}       
\usepackage{url}            
\usepackage{booktabs}       
\usepackage{amsfonts}       
\usepackage{nicefrac}       
\usepackage{microtype}      
\usepackage{xcolor}         
\usepackage{times}
\usepackage{latexsym}

\usepackage{amsmath}
\usepackage{cite}
\usepackage{colortbl}
\usepackage{comment}
\usepackage{adjustbox}
\usepackage{arydshln}
\usepackage{multirow}

\usepackage{pifont}

\usepackage{amssymb}
\usepackage{algorithm}
\usepackage{listings}
\usepackage{placeins}
\usepackage[accsupp]{axessibility}
\usepackage{nccmath}

\usepackage{tcolorbox}

\usepackage{caption} 
\usepackage{subcaption}
\usepackage{makecell}
\usepackage{wrapfig}

\usepackage{mathtools}

\newcommand{\eg}{\textit{e}.\textit{g}.}

\newcommand{\CC}[1]{\cellcolor{#1}}
\definecolor{ftcolor}{rgb}{1,1,1} 
\definecolor{baselinecolor}{rgb}{1,1,1}
\definecolor{contrcolor}{rgb}{1.0, 0.98, 0.8}
\definecolor{predcolor}{rgb}{0.74, 0.83, 0.9}
\definecolor{decorrcolor}{rgb}{0.98, 0.85, 0.87} 
\definecolor{knowcolor}{rgb}{0.80, 0.94, 0.75}
\definecolor{offlinecolor}{rgb}{1,1,1}
\definecolor{firsttaskcolor}{rgb}{0.97, 0.97, 0.97}
\definecolor{azure(colorwheel)}{rgb}{0.0, 0.5, 1.0}
\definecolor{gray(x11gray)}{rgb}{0.75, 0.75, 0.75}
\definecolor{lightgray}{rgb}{0.90, 0.90, 0.90}
\definecolor{darkgray}{rgb}{0.66, 0.66, 0.66}
\definecolor{teagreen}{rgb}{0.82, 0.94, 0.75}
\definecolor{almondlow}{RGB}{252,239,219} 
\definecolor{almondmiddle}{RGB}{237,225,206} 
\definecolor{almondhigh}{RGB}{224,213,194} 
\definecolor{almondultra}{RGB}{214,204,186} 
\definecolor{greylow}{RGB}{235,235,235} 
\definecolor{greymiddle}{RGB}{211,211,211} 
\definecolor{greyhigh}{RGB}{192,192,192} 
\definecolor{pinksecondbest}{RGB}{252, 241, 241}

\definecolor{pastelred}{RGB}{232, 131, 131}
\definecolor{pastelviolet}{RGB}{234, 229, 246}
\definecolor{champagne}{rgb}{0.97, 0.91, 0.81}
\definecolor{lightblue}{RGB}{225, 241, 255}

\usepackage{inconsolata}

\title{OMNI-AVSR: Towards Unified Multimodal Speech Recognition with Large Language Models}

\name{Umberto Cappellazzo$^{\spadesuit}$ \qquad Xubo Liu$^{\clubsuit}$ \qquad Pingchuan Ma$^{\spadesuit}$ \qquad Stavros Petridis$^{\spadesuit}$ \qquad Maja Pantic$^{\spadesuit}$}
  
  \address{$^{\spadesuit}$ Imperial College London, UK \hspace{0.3cm}
      $^{\clubsuit}$ University of Surrey, UK}

\begin{document}

\maketitle
\begin{abstract}
Large language models (LLMs) have recently achieved impressive results in speech recognition across multiple modalities, including Auditory Speech Recognition (ASR), Visual Speech Recognition (VSR), and Audio-Visual Speech Recognition (AVSR). Despite this progress, current LLM-based approaches typically address each task independently, training separate models that raise computational and deployment resource use while missing potential cross-task synergies. They also rely on fixed-rate token compression, which restricts flexibility in balancing accuracy with efficiency. These limitations highlight the need for a unified framework that can support ASR, VSR, and AVSR while enabling elastic inference. To this end, we present Omni-AVSR, a unified audio-visual LLM that combines efficient multi-granularity training with parameter-efficient adaptation. Specifically, we adapt the matryoshka representation learning paradigm to efficiently train across multiple audio and visual granularities, reducing its inherent training resource use. Furthermore, we explore three LoRA-based strategies for adapting the backbone LLM, balancing shared and task-specific specialization. Experiments on LRS2 and LRS3 show that Omni-AVSR achieves comparable or superior accuracy to state-of-the-art baselines while training a single model at substantially lower training and deployment resource use. The model also remains robust under acoustic noise, and we analyze its scaling behavior as LLM size increases, providing insights into the trade-off between performance and efficiency. Our code is available at: \url{https://github.com/umbertocappellazzo/Omni-AVSR}

\end{abstract}

\section{Introduction}
\label{sec:introduction}

Auditory Speech Recognition (ASR) \cite{amodei2016deep, watanabe2017hybrid, prabhavalkar2023end} often degrades in noisy environments such as crowded areas or subways. To address this limitation, Audio-Visual Speech Recognition (AVSR) \cite{dupont2000audio,potamianos2004audio,petridis2018end} incorporates visual cues, such as lip movements, which remain unaffected by acoustic noise, thereby enhancing recognition robustness and accuracy. Early AVSR methods relied on modality-specific encoders and handcrafted fusion strategies \cite{noda2015audio,mroueh2015deep,petridis2018audio}. The introduction of Transformers \cite{vaswani2017attention} significantly advanced performance \cite{afouras2018deep,ma2021end}, spurring research into multimodal learning paradigms such as self-supervision \cite{shi2022learning, shi2022robust, haliassos2023jointly}, ASR-to-AVSR distillation \cite{rouditchenko2024whisper, autoavsr}, and cross-modal complementarity \cite{hong2023watch,chen2023leveraging}.

More recently, Multimodal Large Language Models (MLLMs) have demonstrated that integrating modalities such as vision and speech significantly extends the capabilities of LLMs, yielding state-of-the-art results across diverse tasks \cite{bai2025qwen2, yao2024dense, fathullah2024prompting, goel2025audio, wu2025step}.
Building on this progress, several studies have applied LLMs to ASR, Visual Speech Recognition (VSR), and AVSR, with promising results \cite{llama-AVSR, llama-SMoP, yeo2025mms, yeo2025zero, yeo2024visual, llama-MTSK, cappellazzo2025mome}.

However, most existing approaches treat each task \textbf{in isolation}, training \textit{separate} models for ASR, VSR, and AVSR. This not only increases computational cost and complexity but also overlooks potential synergies across tasks. In contrast, studies across multiple domains have demonstrated the feasibility of unified multi-task multimodal LLMs \cite{zhu2024uni, unifiedmllm, wu2024omni, xu2025qwen2, zhang2025stream}. While some attempts have been made to unify ASR, VSR, and AVSR, these either rely on cumbersome student-teacher pseudo-labeling frameworks \cite{haliassos2024unified} or underperform compared to task-specific models \cite{hsu2022u, rouditchenko2024av, torrie2025multiavsr}.

Motivated by these limitations, we introduce \texttt{Omni-AVSR}, a \textit{unified} audio-visual LLM capable of performing ASR, VSR, and AVSR within a \textbf{single framework}. To adapt the backbone LLM to all tasks in a parameter-efficient manner, we propose three \textit{LoRA}-based methods. Furthermore, we adapt and optimize the \textit{matryoshka representation learning} paradigm \cite{kusupati2022matryoshka, cai2024matryoshka, llama-MTSK} for our setting, enabling \textbf{efficient multi-granularity training} while mitigating its inherent computational cost. This allows the number of tokens to be dynamically adjusted at inference according to resource availability and task requirements. To the best of our knowledge, \texttt{Omni-AVSR} is the \textit{first} audio-visual LLM that supports ASR, VSR, and AVSR jointly while enabling elastic inference under a single set of weights.

Our contributions are summarized as follows:
\textbf{(1)} We provide a comprehensive evaluation of \texttt{Omni-AVSR} on the LRS2 and LRS3 benchmarks, showing that it achieves comparable or superior WER results across all three tasks. Unlike prior methods that support only joint ASR–VSR–AVSR within a single model, only multi-granularity, or neither, Omni-AVSR simultaneously supports both within a single framework, substantially reducing training and deployment resource use. \textbf{(2)} We demonstrate that \texttt{Omni-AVSR} remains competitive with state-of-the-art methods under both clean and noisy conditions. \textbf{(3)} We conduct scaling experiments to analyze the trade-off between LLM size, performance, and computational efficiency.

\section{Omni-AVSR}
\label{sec:method}

\begin{figure*}[t]
    \centering
    \includegraphics[width=0.90\textwidth]{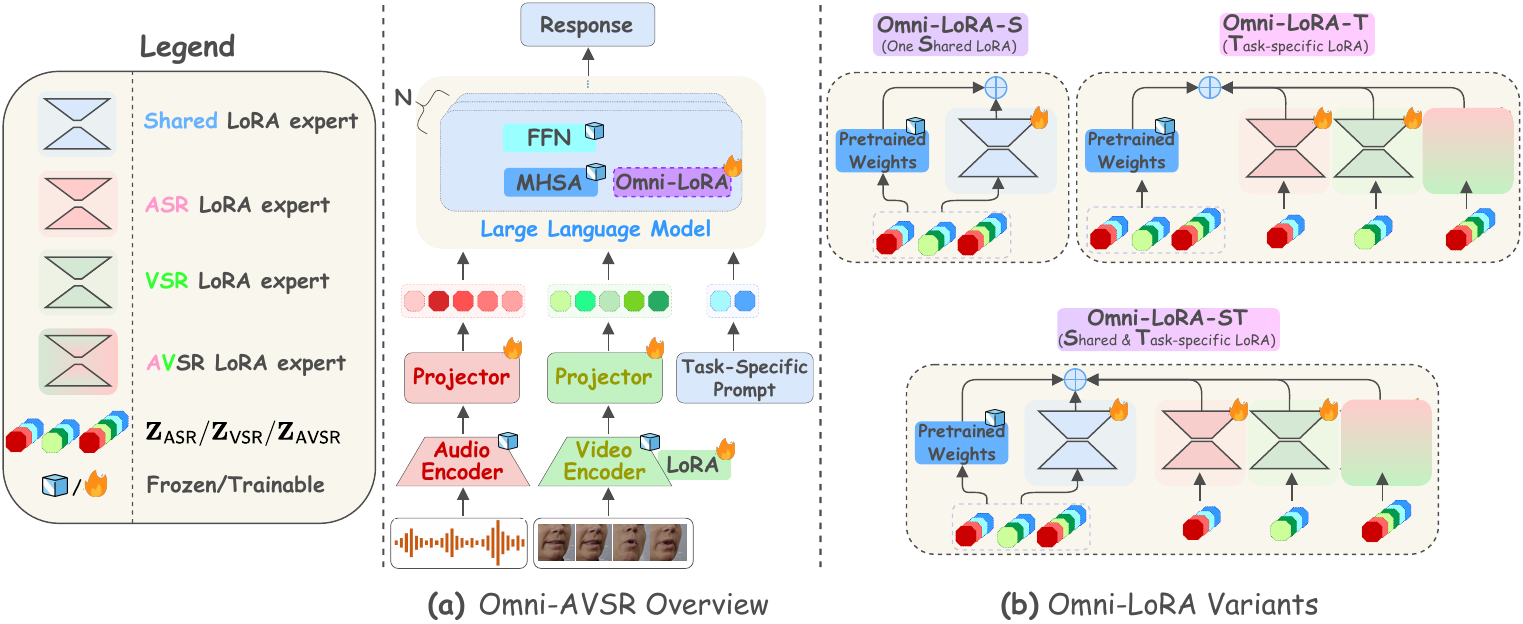}
    \caption{Overview of \textbf{(a)} the proposed \texttt{Omni-AVSR} model and \textbf{(b)} its Omni-LoRA variants. Audio and video inputs are encoded by pre-trained modality-specific encoders and compressed by applying selected audio and video rates before projection into the LLM space. \texttt{Omni-AVSR} explores three LoRA-based LLM adaptation strategies: 1) \textit{Omni-LoRA-S} defines a single LoRA module for both ASR, VSR, and AVSR; 2) \textit{Omni-LoRA-T} dedicates task-specific LoRAs; 3) \textit{Omni-LoRA-ST} makes use of both a shared LoRA and task-specific LoRA modules.}
    \label{fig:main_diagram}
    \vspace{-0.3cm}
\end{figure*}

The goal of \texttt{Omni-AVSR} is to train a single unified LLM-based model capable of performing ASR, VSR, and AVSR. At the same time, it enables flexible control of audio–visual granularity at inference according to resource constraints. In this way, \texttt{Omni-AVSR} supports multiple modalities and granularities within a single set of weights, while reducing training and deployment resource use and achieving performance on par with, or even surpassing, state-of-the-art models trained independently for specific tasks or granularities.

Following prior audio-visual LLMs \cite{llama-AVSR, llama-MTSK, llama-SMoP, yeo2025mms}, \texttt{Omni-AVSR} comprises pre-trained audio and video encoders, projection layers, and an LLM backbone (see Fig. \ref{fig:main_diagram}a). In the next sections, we detail how \texttt{Omni-AVSR} is endowed with \textbf{1)} explicit control over audio-visual granularities during inference and \textbf{2)} the ability to jointly support ASR, VSR, and AVSR within a single model.

\subsection{Multi-Granularity via Efficient Matryoshka Training}

Given an audio waveform $\mathbf{a}$ and its corresponding lip movement video $\mathbf{v}$, we process them with a pre-trained audio encoder (\eg, Whisper \cite{radford2023robust}) and video encoder (\eg, AV-HuBERT \cite{shi2022learning}) to obtain audio and visual tokens, $\mathbf{Z}^a$ and $\mathbf{Z}^v$, respectively. Reducing token granularity lowers computational cost and improves efficiency when feeding audio-visual tokens into an LLM. In AVSR, temporal continuity across modalities creates redundancy, yet most compression methods rely on fixed rates, limiting adaptability to performance and resource trade-offs \cite{llama-AVSR, llama-SMoP, yeo2025mms}. While finer-grained tokens enhance recognition accuracy, they substantially increase inference compute cost due to the quadratic complexity of Transformers.

To address this, Llama-MTSK \cite{llama-MTSK} exploits the matryoshka representation learning (MRL) principle \cite{cai2024matryoshka} to flexibly control audio-visual granularity at inference based on user requirements. During training, token sequences at varying granularities are generated by applying $C_A$ audio compression rates \{$a_1, a_2,\cdots,a_{C_A}$\} and $C_V$ video rates $\{v_1, v_2,\cdots,v_{C_V}$ to the input streams. For AVSR, each of the $C_A\cdot C_V$ audio-visual sequences is fed to the LLM, requiring $C_A\cdot C_V$ forward/backward passes per batch. 

However, when extended to \texttt{Omni-AVSR}, which also supports ASR and VSR, the compute cost grows further: $C_A$ passes for ASR, $C_V$ for VSR, and $C_A\cdot C_V$ for AVSR. \textit{This leads to prohibitive computational overhead and potential interference among multiple objectives}. To overcome this limitation, we introduce a \textbf{key modification}: during training, we randomly select one audio rate $a_i$ and one video rate $v_j$ at each iteration, yielding compressed sequences $\mathbf{Z}^{a_i}$ and $\mathbf{Z}^{v_j}$. This reduces the number of forward/backward LLM passes to only three, one per task, instead of $C_A+C_V+ C_A\cdot C_V$. These compressed sequences are then passed through modality-specific projection layers and concatenated with task-specific text tokens $X^\text{P}_t$, where $t \in \{\mathsf{ASR}, \mathsf{VSR}, \mathsf{AVSR}\}$ and $X^\text{P}_t$ encodes both the task prompt and the transcription. Therefore, we obtain: $\mathbf{Z}_{\mathsf{ASR}} = [\mathbf{Z}^{a_i}, X^\text{P}_{\mathsf{ASR}}]$, $\mathbf{Z}_{\mathsf{VSR}} = [\mathbf{Z}^{v_j}, X^\text{P}_{\mathsf{VSR}}]$, and $\mathbf{Z}_{\mathsf{AVSR}} = [\mathbf{Z}^{a_i}, \mathbf{Z}^{v_j} , X^\text{P}_{\mathsf{AVSR}}]$. This strategy preserves the flexibility of MRL at inference while substantially reducing its training resource use. 

\subsection{Joint ASR-VSR-AVSR Training Formulation}

\texttt{Omni-AVSR} is trained by averaging the auto-regressive next token prediction loss for each task for each input data. The LLM predicts the response $\mathbf{Y} = \{y_s\}_{s=1}^{S}$ conditioned on the multimodal input tokens, where $S$ represents the number of tokens of the ground truth transcription. Accordingly, for each task-specific sequence $\mathbf{Z}_t$, the probability of the target $\mathbf{Y}$ is computed by $p(\mathbf{Y}|\mathbf{Z}_t) = \prod_{s=1}^{S}p_\theta(y_s|\mathbf{Z}_t, y_{<s})$, and the corresponding loss is defined as $\mathcal{L}_t = - \log p(\mathbf{Y}|\mathbf{Z}_t)$, where $y_{<s}$ is the generated output sequence up to token $s-1$, $\theta$ is the trainable parameters, and $t \in \{\mathsf{ASR}, \mathsf{VSR}, \mathsf{AVSR}\}$. Overall, the final objective we train on is:

\begin{equation}
    \mathcal{L}_{\text{OMNI}} = \lambda_{\mathsf{ASR}}\mathcal{L}_{\mathsf{ASR}} + \lambda_{\mathsf{VSR}}\mathcal{L}_{\mathsf{VSR}} + \lambda_{\mathsf{AVSR}}\mathcal{L}_{\mathsf{AVSR}},
    \label{eq:main}
\end{equation}
where $\lambda_{\mathsf{ASR}}$, $\lambda_{\mathsf{VSR}}$, $\lambda_{\mathsf{AVSR}}$ are task-specific weights.

\subsection{Efficient LLM Adaptation via Omni-LoRA}

In \texttt{Omni-AVSR}, following prior works \cite{llama-AVSR, llama-SMoP, llama-MTSK, yeo2024visual, yeo2025mms}, the pre-trained LLM is kept frozen while low-rank LoRA modules \cite{hu2022lora} are employed to parameter-efficiently fine-tune it. Given our multi-task setting, we explore three configurations: \textbf{1)} Omni-LoRA-\textbf{S}, \textbf{2)} Omni-LoRA-\textbf{T}, and \textbf{3)} Omni-LoRA-\textbf{ST}, illustrated in Fig. \ref{fig:main_diagram}b. These variants allow us to systematically investigate the trade-off between parameter sharing and task specialization within \texttt{Omni-AVSR}.

The Omni-LoRA-\textbf{S} variant employs a \textit{single} \textbf{S}hared LoRA module to adapt the query and value projection matrices of each LLM self-attention layer across ASR, VSR, and AVSR tasks. Specifically, a frozen pre-trained weight matrix $W$ is decomposed into low-rank factors with down-projection parameters $W_{down} \in \mathbb{R}^{d \times r}$ and up-projection parameters $W_{up} \in \mathbb{R}^{r \times d}$, where $r \ll d$. Given an input $\mathbf{Z}_t$ for task $t$, the output is computed as: $\mathbf{O}_t = \mathbf{Z}_tW + \alpha(\mathbf{Z}_tW_{down})W_{up}$, where $\alpha$ is a scaling hyperparameter.

The Omni-LoRA-\textbf{T} variant instead defines \textit{separate} \textbf{T}ask-specific LoRA modules, with parameters $W_{down}^t$ and $W_{up}^t$ specialized to each task. The output is then computed as: $\mathbf{O}_t = \mathbf{Z}_tW + \alpha(\mathbf{Z}_tW_{down}^t)W_{up}^t$. Finally, Omni-LoRA-\textbf{ST} combines both \textbf{S}hared \textit{and} \textbf{T}ask-specific LoRA modules, yielding: $\mathbf{O}_t = \mathbf{Z}_tW + \alpha(\mathbf{Z}_tW_{down})W_{up} + \alpha(\mathbf{Z}_tW_{down}^t)W_{up}^t$. During \textbf{training}, Omni-LoRA-T and Omni-LoRA-ST activate all task-specific modules. At \textbf{inference}, however, \textit{only} the module corresponding to the selected task is used, ensuring efficiency.

\section{Experiments and Results}
\label{sec:experiments}

\begin{table}[t]
\centering
    \caption{ASR, VSR, AVSR results in terms of WER (\%) across different audio and video compression rates (e.g., (\texttt{4},\texttt{2})). The best results for each specific task, rate and dataset are shown in \textbf{bold}.}
\resizebox{0.999\linewidth}{!}{
\begin{tabular}{lccccccccc}
\toprule
\multirow{2}{*}{\textbf{Method}} & \multicolumn{2}{c}{\cellcolor{teagreen}\textbf{ASR}} & \multicolumn{2}{c}{\cellcolor{pinksecondbest}\textbf{VSR}} & \multicolumn{4}{c}{\cellcolor{lightblue}\textbf{AVSR}} & \multirow{2}{*}{\textbf{Avg}} \\ \cmidrule(l){2-3}  \cmidrule(l){4-5} \cmidrule(l){6-9}  & (\texttt{4}) & (\texttt{16}) & (\texttt{2}) & (\texttt{5}) & (\texttt{4,2}) & (\texttt{4,5}) & (\texttt{16,2}) & (\texttt{16,5}) \\

\midrule

\multicolumn{10}{c}{\CC{greylow} \textbf{LRS2 Dataset}} \\
Llama-AVSR \cite{llama-AVSR} & 3.3 & 4.3 & 26.9 & 30.0 & \textbf{2.5} & 2.6 & 3.9 & 4.6 & 9.8 \\
Llama-MTSK \cite{llama-MTSK} &\textbf{2.5} &\textbf{3.9}& \textbf{26.7} &28.5 &\textbf{2.5} &2.5 &3.7 &4.0 &\textbf{9.3} \\
Llama-MT &2.6 &4.1 &27.2 &28.8 &\textbf{2.5} &\textbf{2.4} &\textbf{3.5} &\textbf{3.9} &9.4 \\
\hdashline \addlinespace[2pt]

\rowcolor{pastelviolet}\textbf{\texttt{Omni-AVSR-S}} &2.8 &5.0 &27.8  &28.5 &2.7 &2.6 &3.8 &4.0 &9.6 \\
\rowcolor{pastelviolet}\textbf{\texttt{Omni-AVSR-T}} &2.7 &4.5 &26.8 &\textbf{28.3} &2.6 &2.7 &3.9 &4.0 &9.4 \\
\rowcolor{pastelviolet}\textbf{\texttt{Omni-AVSR-ST}} &2.7 &4.8 &27.8 &29.5 &\textbf{2.5} &2.7 &3.9 &4.2 &9.8 \\\addlinespace[2pt]

\multicolumn{10}{c}{\CC{greymiddle} \textbf{LRS3 Dataset}} \\
Llama-AVSR \cite{llama-AVSR} &1.1 &2.0 &27.4 &29.5 &1.1 &1.2 &2.0 &2.1 &8.3 \\
Llama-MTSK \cite{llama-MTSK} &\textbf{1.0} &2.0 &26.9 &27.8 &\textbf{1.0} &\textbf{1.0} &1.9 &2.0 &8.0 \\
Llama-MT &\textbf{1.0} &2.1 &27.2 &28.4 &\textbf{1.0} &\textbf{1.0} &\textbf{1.8} &\textbf{1.9} &8.0 \\
\hdashline \addlinespace[2pt]
\rowcolor{pastelviolet}\textbf{\texttt{Omni-AVSR-S}} &1.1 &2.4 &\textbf{26.6} &27.4 &1.1 &\textbf{1.0} &1.9 &2.0 &\textbf{7.9} \\
\rowcolor{pastelviolet}\textbf{\texttt{Omni-AVSR-T}} &1.2 &\textbf{1.9} &26.7 &27.8 &1.2 &1.2 &2.0 &2.2 &8.0  \\
\rowcolor{pastelviolet}\textbf{\texttt{Omni-AVSR-ST}} &1.2 &2.0 &26.8 &\textbf{27.1} &\textbf{1.0} &1.1 &\textbf{1.8} &\textbf{1.9} &\textbf{7.9} \\

\bottomrule

\end{tabular}}
\label{tab:main}
\vspace{-0.2cm}
\end{table}

\begin{table}[t]
\renewcommand{\tabcolsep}{0.9mm}
\centering
    \caption{Computational cost analysis in terms of \textbf{1)} the number of trained models and \textbf{2)} LLM forward/backward passes required to cover all tasks and rates in training. Here, $T$ denotes the number of tasks, while $C_A$/$C_V$ denotes the number of \textit{audio}/\textit{video} rates.}
\begin{tabular}{lcc}

\toprule
\textbf{Method} & \textbf{\# Trained Models} & \textbf{\# LLM F/B Passes} \\

\midrule

\textbf{Llama-AVSR} \cite{llama-AVSR} & $C_A$ + $C_V$ + $C_A C_V$ & $C_A$ + $C_V$ + $C_A C_V$ \\
\textbf{Llama-MTSK} \cite{llama-MTSK} & $T$ & $C_A$ + $C_V$ + $C_A C_V$ \\
\textbf{Llama-MT} & $C_A C_V$ & $T$($C_AC_V$) \\
\hdashline \addlinespace[2pt]
\rowcolor{pastelviolet}\textbf{\texttt{Omni-AVSR}} & $\textbf{1}$ & $\textbf{T}$ \\

\bottomrule
 \end{tabular}
\label{tab:computation_cost}
\vspace{-0.3cm}
\end{table}

\subsection{Experiment Settings}
\textbf{Datasets}. We conduct experiments on LRS2 \cite{chung2017lip} and LRS3 \cite{afouras2018lrs3} datasets. LRS2 includes $225$ hours of footage from BBC programs. LRS3 contains $433$ hours of English video clips from TED talks.  

\textbf{Pre-Processing}. We follow \cite{autoavsr, llama-AVSR, llama-MTSK} for the datasets pre-processing. For video, we crop the mouth region of interests (ROIs) through a bounding box of $96$ × $96$. Each frame is normalised by subtracting the mean and dividing by the standard deviation of the training set. Audio data undergo z-normalisation per utterance.

\textbf{\texttt{Omni-AVSR} Details}. We use AV-HuBERT Large as the visual encoder and Whisper medium as the audio encoder. The projection layers consist of two linear layers with a ReLU activation in between. For the LLM backbone, we adopt LLaMA 3.2-1B \cite{dubey2024llama} in our main experiments. Following prior work \cite{llama-AVSR, llama-MTSK, llama-SMoP}, both the LLM and video encoder are fine-tuned via LoRA modules applied to the query and value projection matrices with rank $64$. 

\textbf{Training/Inference Details}. Following \cite{autoavsr, llama-AVSR, llama-MTSK}, we augment visual inputs through horizontal flipping, random cropping, and adaptive time masking, while for audio we apply adaptive time masking. 
We set the textual prompts as in \cite{llama-AVSR, llama-MTSK, llama-SMoP}: ``\texttt{Transcribe \{\textbf{task\_prompt}\} to text.}'', where \texttt{\textbf{task\_prompt}} $\in$ \{``\texttt{speech}'', ``\texttt{video}'', ``\texttt{speech and video}''\}. We set $\lambda_\mathsf{ASR} = \lambda_{\mathsf{AVSR}} = 1$ and $\lambda_\mathsf{VSR} = 1.5$. We train our \texttt{Omni-AVSR} models for $8$ epochs with AdamW optimizer with cosine annealing scheduler and weight decay set to $0.1$.  The learning rate is 1e-3. For decoding, we use beam search with a beam width of $15$ and temperature of $0.6$. 

\textbf{Audio-Visual Granularities}. For fair comparison with prior work \cite{llama-MTSK}, we adopt the same compression rates, chosen to capture a spectrum of efficiency–performance trade-offs at inference. Specifically, we use \{$\texttt{4}$,$\texttt{16}$\} for ASR, \{$\texttt{2}$,$\texttt{5}$\} for VSR, and their Cartesian product for AVSR, yielding four audio-visual configurations. Token compression is performed via \textit{average pooling} \cite{llama-MTSK}.

\textbf{Baselines}. As shown in Tables \ref{tab:main} and \ref{tab:noisy}, we compare \texttt{Omni-AVSR} variants with three main approaches: 1) \textbf{Llama-AVSR} \cite{llama-AVSR}, which trains a separate model for each task and compression rate; 2) \textbf{Llama-MTSK} \cite{llama-MTSK}, which enables elastic inference by training on multiple rates but only within a single task; 3) \textbf{Llama-MT}, which supports multi-task (\textbf{MT}) training across ASR, VSR, and AVSR but requires a separate model for each rate. In contrast, \texttt{Omni-AVSR} unifies both elastic inference and multi-task learning within a single framework, subsuming these baselines as special cases. Additional comparisons with AVSR sota methods are provided in Section \ref{sec:ablation}.

\begin{table}[t]
\centering
    \caption{AVSR results on LRS3 across acoustic noise conditions.}
\begin{tabular}{lccccc}

\toprule
\multirow{2}{*}{\textbf{Method}} & \multicolumn{5}{c}{\cellcolor{almondlow}\textbf{SNR (dB)}}\\
      \cmidrule(rl){2-6} & 5 & 2.5 & 0 & -2.5 & -5\\

\midrule

\multicolumn{6}{c}{\CC{greylow} \textbf{Compression rates: (\texttt{4,2})}} \\
Llama-AVSR \cite{llama-AVSR} &2.6 &4.1 &4.8 &12.1 &19.1 \\
Llama-MTSK \cite{llama-MTSK} &2.5 &3.9 &4.8 &11.7 &18.5 \\
Llama-MT &2.6 &3.9 &4.4 &11.1 &17.8\\ \hdashline \addlinespace[2pt]
\rowcolor{pastelviolet}\textbf{\texttt{Omni-AVSR-ST}} &2.5 &3.8 &4.4 &11.4 &18.0 \\ \addlinespace[2pt]

\multicolumn{6}{c}{\CC{greymiddle} \textbf{Compression rates: (\texttt{16,5})}} \\
Llama-AVSR \cite{llama-AVSR} &4.2 &5.8 &6.5 &14.9 &22.1 \\
Llama-MTSK \cite{llama-MTSK} &3.8 &5.5 &6.0 &14.0 &20.5 \\
Llama-MT &3.7 &5.1 &6.0 &13.4 &20.1 \\\hdashline \addlinespace[2pt]
\rowcolor{pastelviolet}\textbf{\texttt{Omni-AVSR-ST}} &3.9 &5.3 &5.9 &13.5 &19.5 \\

\bottomrule
 \end{tabular}
\label{tab:noisy}
\vspace{-0.2cm}
\end{table}

\begin{table}[t]
\renewcommand{\tabcolsep}{1.mm}
\centering
    \caption{Comparison with state-of-the-art methods using a single model for ASR, VSR, and AVSR on LRS3. $^\ddagger$u-HuBERT and AV-CPL are trained on LRS3 and VoxCeleb2, totaling $1759$ hours.}
\begin{tabular}{lccccc}

\toprule
\multirow{2}{*}{\textbf{Method}} & \textbf{Train} & \textbf{Train} & 
\multicolumn{3}{c}{\textbf{WER} $\downarrow$} \\
& \textbf{Par. (M)} & \textbf{Hours} &\cellcolor{teagreen}\textbf{ASR} &\cellcolor{pinksecondbest}\textbf{VSR} &\cellcolor{lightblue}\textbf{AVSR} \\

\midrule

u-HuBERT \cite{hsu2022u}$^\ddagger$ &325 &1759 &1.5 &29.1 &1.3 \\
AV-CPL \cite{rouditchenko2024av}$^\ddagger$ &325 & 1759 & 2.3 & 47.4 & 2.2 \\
MultiAVSR \cite{torrie2025multiavsr} &274 &433 &2.4 &31.1 &2.5 \\
USR \cite{haliassos2024unified} &171 &433 &1.9 &34.3 &1.6 \\ \hdashline \addlinespace[2pt]
\rowcolor{pastelviolet}\textbf{\texttt{Omni-AVSR-ST}} (\texttt{4,2})&\textbf{58} &433 &\textbf{1.2} &\textbf{26.8} &\textbf{1.0} \\
\rowcolor{pastelviolet}\textbf{\texttt{Omni-AVSR-ST}} (\texttt{16,5})&\textbf{58}&433 &2.0 &27.1 &1.9 \\

\bottomrule
 \end{tabular}
\label{tab:comparison_unified_AVSR_methods}
\vspace{-0.3cm}
\end{table}

\begin{figure*}
     \centering
     \begin{subfigure}[b]{0.43\textwidth}
         \centering
         \includegraphics[width=\textwidth]{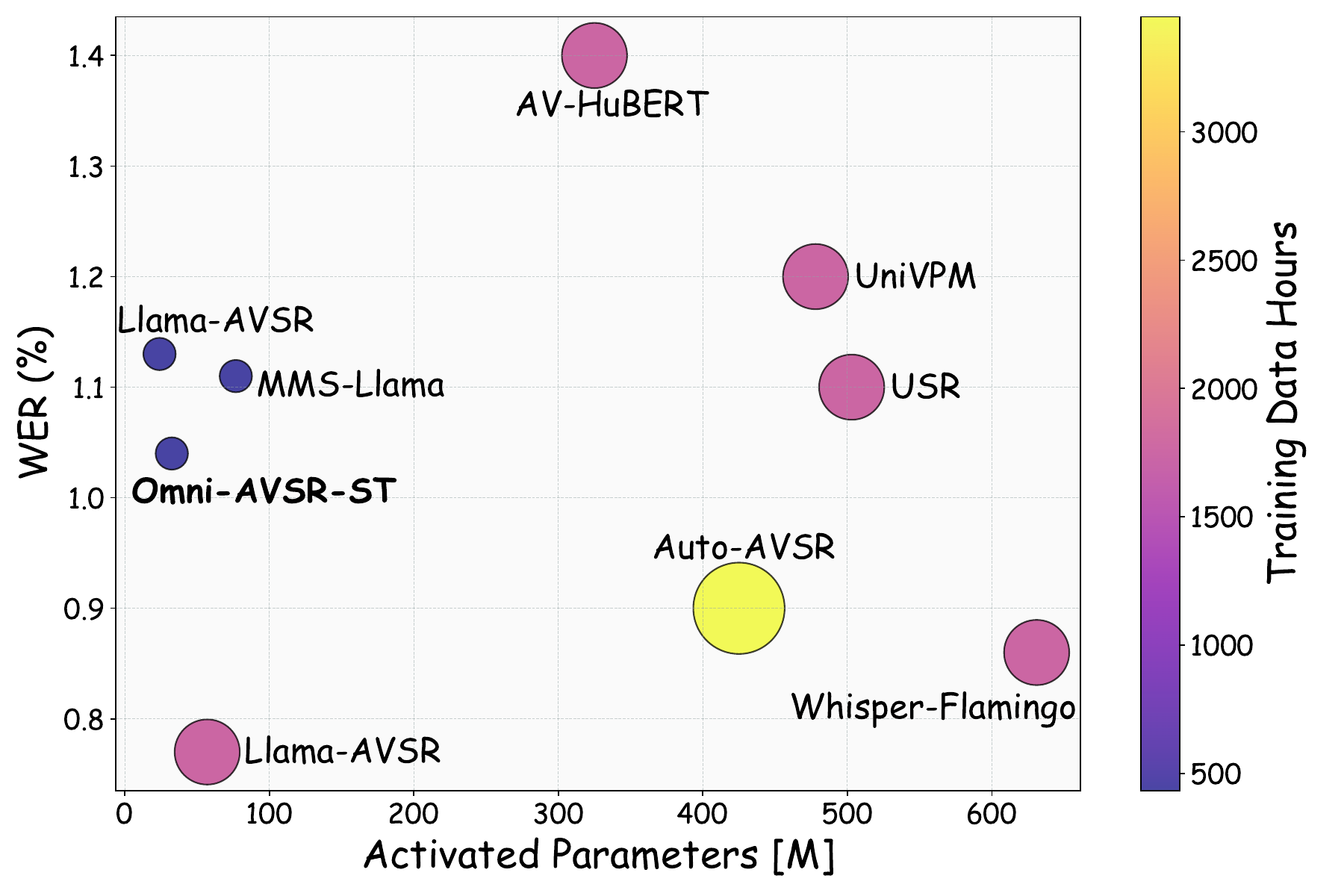}
     \end{subfigure}
     \begin{subfigure}[b]{0.35\textwidth}
         \centering
         \includegraphics[width=\textwidth]{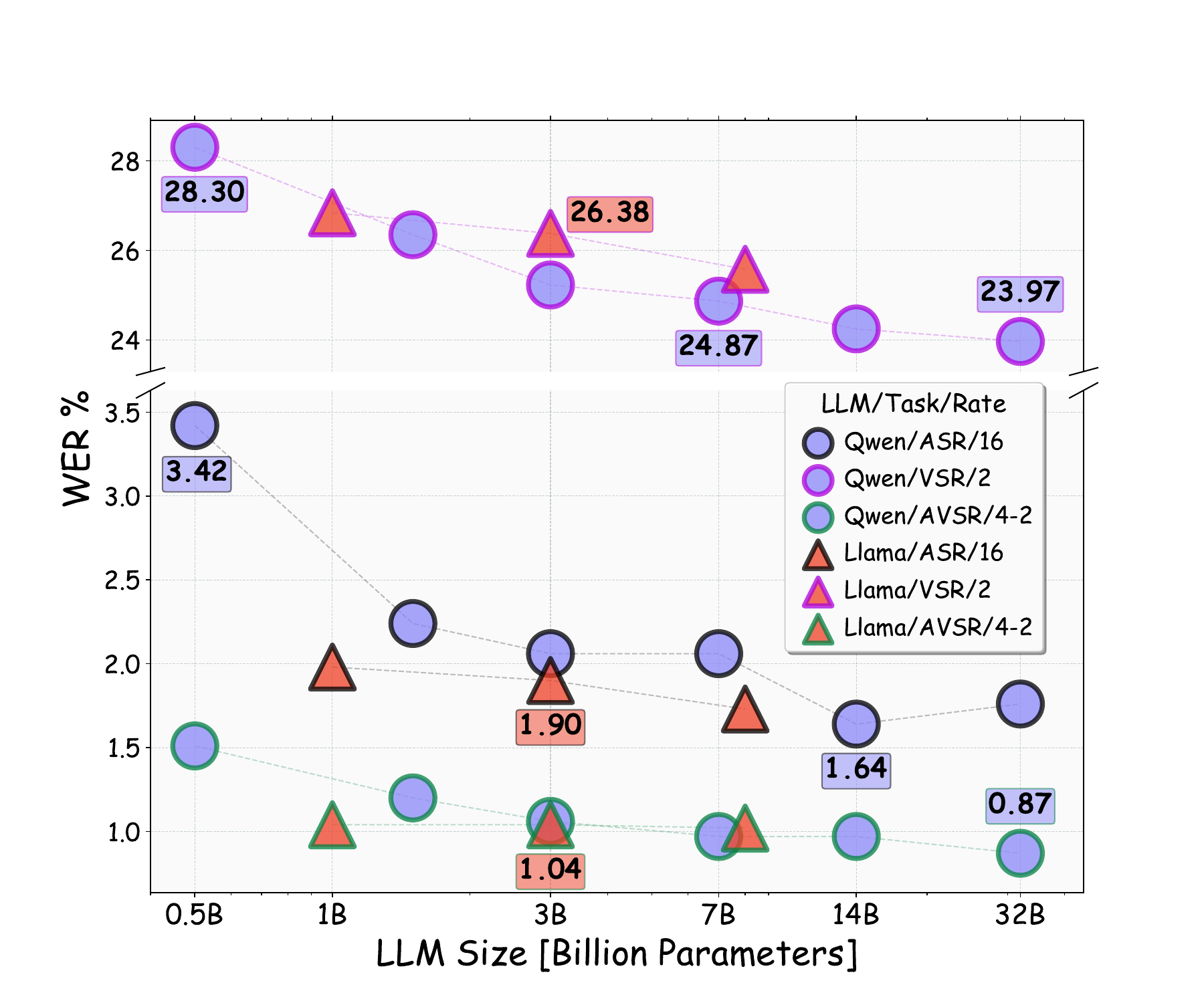}
     \end{subfigure}
    
    \caption{\textbf{Left}: Comparison of \texttt{Omni-AVSR-ST} with state-of-the-art AVSR methods in terms of \textit{WER}, \textit{activated parameters}, and \textit{training data hours} on LRS3. \textbf{Right}: \textit{Scaling} trend of \texttt{Omni-AVSR-ST} when we increase the LLM size on LRS3.}
    \label{fig:bubble_scaling}
    \vspace{-0.5cm}
\end{figure*}

\subsection{Main Results}
Table \ref{tab:main} reports the ASR/VSR/AVSR results of our three \texttt{Omni-AVSR} variants on LRS2 and LRS3. On LRS2, among the three proposed variants, \texttt{Omni-AVSR-T} achieves the best performance, while on LRS3 all three variants yield comparable results. This difference is likely due to the larger training set of LRS3, which enables lower WERs overall, particularly for ASR and AVSR. Compared with the baselines, we observe the following: \textbf{(1)} all \texttt{Omni-AVSR} variants consistently outperform Llama-AVSR, which requires a separate model per rate and task; \textbf{(2)} \texttt{Omni-AVSR-T} on LRS2, and all three variants on LRS3, match or surpass Llama-MTSK and Llama-MT, with \texttt{Omni-AVSR-S} and \texttt{-T} attaining average WERs as low as $7.9$ across tasks on LRS3; \textbf{(3)} task-wise, \texttt{Omni-AVSR} particularly benefits VSR; and \textbf{(4)} performance trends remain consistent across compression rates. These results demonstrate that \texttt{Omni-AVSR} delivers competitive or superior accuracy while unifying elastic inference and multi-task learning within a single framework.

Beyond delivering strong recognition performance, \texttt{Omni-AVSR} also offers significant computational advantages, as summarized in Table \ref{tab:computation_cost}. \textbf{(1)} \texttt{Omni-AVSR} requires training only a \textit{single} model, independent of the number of tasks $T$ (ASR, VSR, and AVSR in our case, so $T=3$) and the number of audio $C_A$ and video $C_V$ compression rates ($C_A = C_V = 2$ in our setup). In contrast, the other three baselines train multiple models that scale with the number of compression rates (Llama-AVSR and Llama-MT) or with the number of tasks (Llama-MT). \textbf{(2)} We further compare the methods in terms of the number of forward/backward passes required over the LLM, since this constitutes the dominant computational cost during training. Llama-AVSR and Llama-MTSK must compute the loss separately for each compression rate and task, requiring $C_A$ passes for ASR, $C_V$ for VSR, and $C_AC_V$ for AVSR. Llama-MT trains one multi-task model for each audio–visual rate pair, which results in $T(C_AC_V)$ passes. In contrast, \texttt{Omni-AVSR} computes the loss only once per task, as it samples a single audio and video rate at each iteration, thus reducing the requirement to just $T$ passes. Overall, \texttt{Omni-AVSR} \textit{requires only a single model and substantially reduces overall training computations compared to all baselines}.

\textbf{Results under Acoustic Noise}. To evaluate the robustness of \texttt{Omni-AVSR} under \textit{noisy conditions}, we inject babble noise from the NOISEX dataset \cite{noisex} at varying SNRs. As shown in Table \ref{tab:noisy}, \texttt{Omni-AVSR-ST} consistently outperforms Llama-AVSR and Llama-MTSK, and remains competitive with Llama-MT across noise levels, often surpassing it at lower SNRs. 

\textbf{Comparison with Other Multi-task Methods}. 

In Table \ref{tab:comparison_unified_AVSR_methods}, we compare \texttt{Omni-AVSR-ST} with three state-of-the-art methods that train a single model for ASR, VSR, and AVSR: u-HuBERT \cite{hsu2022u}, AV-CPL \cite{}, MultiAVSR \cite{torrie2025multiavsr}, and USR \cite{haliassos2024unified}. Unlike \texttt{Omni-AVSR}, these methods do not support elastic inference. At the (\texttt{4},\texttt{2}) compression setting, \texttt{Omni-AVSR-ST} achieves the best performance across all tasks while requiring significantly fewer parameters and surpassing u-HuBERT and AV-CPL, despite both being trained on $1759$ hours of data (LRS3 + VoxCeleb2 datasets). Even under the more extreme (\texttt{16},\texttt{5}) compression, \texttt{Omni-AVSR-ST} maintains competitive results within a single set of weights.

\subsection{Ablation Studies}
\label{sec:ablation}

\begin{table}[t]
\renewcommand{\tabcolsep}{1.5mm}
\centering
    \caption{Ablation on the best values of ASR/VSR/AVSR weights.}
\begin{tabular}{ccccccccc}

\toprule
\multirow{2}{*}{$\lambda_{\mathsf{ASR}}$} & \multirow{2}{*}{$\lambda_\mathsf{VSR}$} & \multirow{2}{*}{$\lambda_{\mathsf{AVSR}}$} & \multicolumn{2}{c}{\cellcolor{teagreen}\textbf{ASR}} & \multicolumn{2}{c}{\cellcolor{pinksecondbest}\textbf{VSR}} &\multicolumn{2}{c}{\cellcolor{lightblue}\textbf{AVSR}} \\ \cmidrule(l){4-5}  \cmidrule(l){6-7} \cmidrule(l){8-9} & & & (\texttt{4}) & (\texttt{16}) & (\texttt{2}) & (\texttt{5}) & (\texttt{4,2}) & (\texttt{16,5}) \\

\midrule

1&1&1 &2.9 &5.7 &27.0&28.6&2.7&4.4 \\
\rowcolor{pastelviolet} 1&1.5&1 &2.7 &4.4 &26.8&28.3&2.6&4.0 \\
1&2&1 &2.7&4.4 &27.0&28.5&2.5&4.0 \\

\bottomrule
 \end{tabular}
\label{tab:ablation_coeff}
\vspace{-0.3cm}
\end{table}

\textbf{AVSR Comparison with Sota Methods}. Fig. \ref{fig:bubble_scaling} (left) presents a comparison of \texttt{Omni-AVSR-ST} with recent state-of-the-art approaches on LRS3 for the AVSR task. Baselines include UniVPM \cite{hu2023hearing}, USR \cite{haliassos2024unified}, Whisper-Flamingo \cite{rouditchenko2024whisper}, Llama-AVSR \cite{llama-AVSR}, Auto-AVSR \cite{autoavsr}, AV-HuBERT \cite{shi2022learning}, and MMS-Llama \cite{yeo2025mms}. \texttt{Omni-AVSR-ST} (evaluated at audio-video rates of (\texttt{4},\texttt{2})) achieves competitive WERs while requiring substantially fewer parameters and training data hours than all baselines, within one consistent framework.

\textbf{LLM Scaling Trend}. We study how scaling the LLM impacts performance in Fig. \ref{fig:bubble_scaling} (right). Specifically, we evaluate Llama 3.2–1B, 3B, and 3.1–8B \cite{dubey2024llama}, as well as Qwen 2.5–0.5B, 1.5B, 3B, 7B, 14B, and 32B \cite{qwen2.5}. Results are reported for ASR at audio rate $\texttt{16}$ (black outline), VSR at video rate \texttt{2} (violet outline), and AVSR at (\texttt{4},\texttt{2}) rates. As shown, performance improves with larger LLMs, with higher gains observed on more challenging tasks (e.g., VSR) or under higher compression (e.g., ASR at rate \texttt{16}). However, larger models incur greater training computations, memory usage, and slower inference. Overall, LLMs in the 1–3B parameter range represent a favorable trade-off between accuracy and efficiency.

\textbf{Optimal Task-specific Weights}. In Table \ref{tab:ablation_coeff}, we analyze the impact of varying the loss weight coefficients in Eq. \ref{eq:main} for each task on LRS2. The best performance is obtained with $\lambda_{\mathsf{ASR}} = \lambda_{\mathsf{AVSR}} = 1$ and $\lambda_{\mathsf{VSR}} = 1.5$. Since VSR is the most challenging of the three tasks, assigning it a higher weight leads to improved overall results.

\section{Conclusion}
\label{sec:conclusion}

In this work, we present \texttt{Omni-AVSR}, the first unified audio-visual LLM that jointly supports ASR, VSR, and AVSR while enabling elastic inference under a single set of weights. By combining efficient matryoshka-based multi-granularity training with LoRA adaptation strategies, \texttt{Omni-AVSR} achieves strong performance while reducing training and deployment resource use. Experiments on LRS2 and LRS3 show that \texttt{Omni-AVSR} matches or surpasses state-of-the-art baselines, remains robust in noisy conditions, and delivers favorable trade-offs when scaling LLM size. Furthermore, \texttt{Omni-AVSR} provides significant computational savings, requiring only one model and a reduced number of LLM passes during training. 


\clearpage
\bibliographystyle{IEEEtran}
\bibliography{refs}

\end{document}